# Feature Selection via Block-Regularized Regression

**Seyoung Kim**
School of Computer Science
Carnegie Mellon University
Pittsburgh, PA 15213

**Eric Xing**
School of Computer Science
Carnegie Mellon University
Pittsburgh, PA 15213

## Abstract

Identifying co-varying causal elements in very high dimensional feature space with internal structures, e.g., a space with as many as millions of linearly ordered features, as one typically encounters in problems such as whole genome association (WGA) mapping, remains an open problem in statistical learning. We propose a block-regularized regression model for sparse variable selection in a high-dimensional space where the covariates are linearly ordered, and are possibly subject to local statistical linkages (e.g., block structures) due to spacial or temporal proximity of the features. Our goal is to identify a small subset of relevant covariates that are not merely from random positions in the ordering, but grouped as contiguous blocks from large number of ordered covariates. Following a typical linear regression framework between the features and the response, our proposed model employs a sparsity-enforcing Laplacian prior for the regression coefficients, augmented by a 1st-order Markovian process along the feature sequence that "activates" the regression coefficients in a coupled fashion. We describe a sampling-based learning algorithm and demonstrate the performance of our method on simulated and biological data for marker identification under WGA.

## 1 INTRODUCTION

Recent advances in high-throughput genotyping technology have allowed researchers to generate a high volume of genotype data at a relatively low cost. An association study involves examining genotype data of individuals in a population and phenotype data for the same individuals such as disease status, gene ex-pression, and physiological measurements in order to discover genetic markers that affect the phenotype of the individual possessing a particular variation of the marker. Variations in a single nucleotide called single nucleotide polymorphisms (SNPs) provide a useful set of genetic markers since they are relatively common across the genome. The whole-genome association study has become feasible because of the relatively low cost involved in typing a large number of SNPs. The challenge in this type of study is to identify a small subset of SNPs associated with the phenotype among the full set of SNPs that can be as large as 8 million (the International HapMap Consortium, 2005).

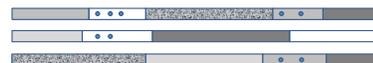

Figure 1: An illustration of linkage disequilibrium in chromosomes in an association study.

A simple single-marker test has been widely used for detecting an association (Stranger et al. 2005, Cheung et al. 2005). Using this approach, one examines the correlation between the given phenotype and frequencies of each polymorphic allele of one SNP marker at a time to compute $p$-value of the SNP, and finds the SNPs with low $p$-values to be significant. The single-marker test assumes that SNPs are independent of each other, ignoring an important correlation structure due to the *linkage disequilibrium* present in the sequence of SNPs. In reality, the states of SNPs that are adjacent in the genome can be tightly coupled (i.e., in linkage disequilibrium). This is because when an individual inherits a chromosomal material from each of the parents, a recombination event can break the parental chromosomes into non-random inheritable segment, causing SNPs within the segment to be inherited with high probability, and preventing random combinations of all possible SNP states within the segment. Since the recombination sites are non-uniformly distributed across the genome, recombination events in chromosomes in a population over generations lead to a block structure in SNPs on the chromosome. As illustrated in Figure 1, each chromosome is a mosaic of ancestor chromosomes, where segments of SNPs of the same color have been inherited from the same ancestor chromosome. The true association SNPs called causal SNPs are indicated as circles in Figure 1. Since a chromosome segment carrying causal alleles can be inherited as a block, we can take advantage of this block structure to increase the power of the study for detecting association by considering a block of linked SNPs jointly rather than a single SNP at a time.

A multi-marker approach takes into account this linkage disequilibrium pattern by testing a short segment of SNP markers called a haplotype for an association (Zailen et al. 2007, Zhang et al. 2002). In this case, a haplotype instead of a single SNP acts as a proxy for untyped causal SNPs. However, they test a haplotype of a fixed length for an association, scanning the genome using a sliding window. Most of these approaches do not explicitly make use of the block structure with possibly varying block lengths in the sequence of SNPs.

In this paper, we propose a model for association mapping that explicitly incorporates the linkage disequilibrium pattern. We focus on continuous valued phenotypes, and base our method on a linear regression model, where the SNPs are predictors and the phenotypes are response variables. The SNPs with large regression coefficients are found to be significant. The number of SNPs involved in a typical association study is very large, and we are interested in extracting a small number of causal SNPs that are grouped into blocks due to linkage disequilibrium. Thus, we can view this problem as 1) identifying relevant covariates when the covariates lie in a high-dimensional space and 2) learning a block structure in those relevant covariates at the same time.

To enforce sparsity in the regression model, we use the Laplacian prior on the regression coefficients, similar to the $L_1$ penalty in the lasso (Tibshirani 1996). However, the lasso does not provide any mechanism to incorporate the correlation structure in covariates into the model to address the second problem. In this paper, we propose to encode the information of the correlation pattern in the SNPs as a Markov chain, and use this Markov chain as a prior in the regression model. The block boundaries for chromosome regions with a high level of association are determined probabilistically through transition probabilities in the Markov chain. A regression-based association has been used previously (Servin and Stephens 2007), but they did not address the problem of taking into account the block structure in the genome in a high dimensional space to improve the power of the study.

There is a large body of literature on variable selection methods such as the lasso (Tibshirani 1996) and Bayesian variable selection algorithms (George and McCulloch 1993, Ishwaran and Rao 2005, Yuan and Lin, 2005). Most of these works did not consider situations in which the covariates are structured in a certain manner. Nott and Green (2004) used the correlation information in the covariates to improve the convergence of sampling algorithm, but the model itself did not assume any structure in the covariates. In the fused lasso (Tibshirani et al. 2005), covariates were assumed to be ordered, and adjacent regression coefficients tended to be fused to take the same value, encouraging sparsity in the difference between adjacent coefficients. However, it used only the ordering information, and did not take into account the additional information on the structure in covariates such as the linkage disequilibrium pattern in the case of association study. In the group lasso (Yuan and Lin 2005), the group structure in covariates was assumed to be known, whereas in our proposed model we determine the block structure of relevant covariates during learning given prior knowledge on the correlation structure.

The rest of the paper is organized as follows. In Section 2, we describe the proposed model for association mapping and the learning algorithm. In Section 3, we apply the model to simulated data and mouse data, and compare the performance of the proposed model with that of existing methods. In Section 4, we conclude with a brief discussion of future work.

## 2 GENOME-WIDE ASSOCIATION VIA BLOCK-REGULARIZED REGRESSION

### 2.1 THE MODEL

**The Association Model** Let us assume that a set of $J$ SNP markers have been typed for $N$ individuals. The genotype of each SNP in an individual consists of two alleles corresponding to each of a pair of chromosomes in the case of diploid organisms such as humans and mice. For our regression analysis, we construct an $N \times J$ design matrix $\mathbf{X}$, where each element $x_{ij}$ takes values from $\{0, 1, 2\}$ according to the number of minor alleles in the genotype of the $j$th SNP of the $i$th individual, assuming a strictly additive genetic effect of the minor allele. Note that in our analysis the $J$ SNP markers are assumed to be ordered in terms of their positions on chromosome. In addition, a measurement of phenotype $y_i$ is available for each individual $i$, and we let $\mathbf{y}$ be an $N \times 1$ vector of such measurements for the $N$ individuals. In particular, the covariates in $\mathbf{X}$ lie in a high dimensional space with only a small sub-set of the $J$ SNPs influencing the output $\mathbf{y}$. In this setting, we assume a standard linear regression model as follows:

$$\mathbf{y} = \mathbf{X}\boldsymbol{\beta} + \epsilon, \quad \epsilon \sim N(0, \sigma^2),$$

where $\boldsymbol{\beta}$ is a vector of $J$ regression coefficients $\{\beta_1, \ldots, \beta_J\}$, and the noise $\epsilon$ is modeled as having a normal distribution with mean 0 and variance $\sigma^2$. Taking a Bayesian approach, we set the prior for $\sigma^2$ to Inv-gamma$(\nu_0/2, (\nu_0 s_0^2)/2)$.

We use a prior on $\boldsymbol{\beta}$ that enforces sparsity in the coefficients. We model each regression coefficient $\beta_j$ as coming from a mixture of two components, one component representing irrelevant covariates (non-causal SNPs) and the other for relevant covariates (causal SNPs). We introduce a random variable $c_j$ that takes values from $\{0, 1\}$ to indicate the mixture component label for $\beta_j$. If $c_j = 0$, the $j$th SNP is non-causal, and $\beta_j$ is set to 0. If $c_j = 1$, then we model $\beta_j$ as coming from a Laplacian distribution. The complete probability distribution for $\beta_j$ given $c_j$ is given as

$$\beta_j | c_j \sim \begin{cases} I(\beta_j = 0) & \text{if } c_j = 0 \\ \frac{1}{2(2\lambda\sigma^2)} \exp\left(-\frac{|\beta_j|}{2\lambda\sigma 2}\right) & \text{if } c_j = 1, \end{cases} \quad (1)$$

where $\lambda$ is the parameter that controls the amount of sparsity. We use Inv-gamma$(\alpha, \gamma)$ as a prior distribution on $\lambda$, and sample $\lambda$ from its posterior during learning (Park and Casella 2008).

The Laplacian prior in Equation (1) is the same as the $L_1$ penalty used in the lasso (Tibshirani 1996), and has been shown to be useful in a more general problem of learning a sparse model in high-dimensional space (Wainwright et al. 2006). In a Bayesian setting, models similar to the one described above have been used in the literature of Bayesian variable selection with various different distributions in Equation (1) (George and McCulloch 1993, Ishwaran and Rao, 2005, Yuan and Lin, 2005). In almost all of these works, $c_j$'s were modeled as coming from a Bernoulli distribution with parameter $p$, assuming that $c_j$'s are independent of each other. This independence assumption is not appropriate in the case of genetic data since nearby SNP markers are known to be highly correlated due to the linkage disequilibrium. In the next section, we propose to use a Markov chain prior for $c_j$'s that takes advantage of this dependency.

**Markov Chain Prior for Block Structure** Because of the linkage disequilibrium, individual chromosomes tend to have a block structure in the sequence of genetic markers, and such blocks of genetic markers are shared across individuals in a population. Recombination rate summarizes the degree of correlation between tightly linked SNPs. In a region with a high recombination rate, the previously linked SNPs are likely to be decoupled, resulting in a weak correlation, whereas a segment of tightly linked SNPs is preserved in the absence of recombination during inheritance. Becuase of the linkage disequilibrium structure, considering a block of highly correlated SNPs instead of a single SNP for an association with the phenotype can potentially increase the power of the association study for detecting causal SNPs.

In this section, we propose to use a Markov chain prior on the indicator variable $\mathbf{c}$ to take into account the block structure due to the linkage disequilibrium in SNP markers by incorporating the recombination rate information in the prior. There is a large body of literature on estimating recombination rates given genotype data of unrelated individuals from a population (Fearnhead and Donnelly 2001, Li and Stephens 2003, Sohn and Xing, 2007). Any of these methods can be used to estimate recombination rates as part of a preprocessing step prior to the association analysis through the proposed method.

Given the estimated recombination rates, we assume that the $J$ covariates are ordered in their positions to have a chain structure and that there is an implicit block structure in the chain where the block boundaries are defined stochastically in terms of the distance and recombination rate between each pair of covariates. In this setting, a group of SNPs within a block can be assigned together to be either causal or non-causal.

We model the sequence of indicator variables $\mathbf{c} = \{c_1, \ldots, c_J\}$ as a Markov chain as follows:

$$P(\mathbf{c}) = P(c_1) \prod_{j=2}^{J} P(c_j | c_{j-1}).$$

For $P(c_j | c_{j-1})$, we use a Poisson process model, a model commonly used for recombination process (Li and Stephens 2003),

$$P(c_j | c_{j-1}) = \exp(-d_j \rho_j) \, \delta(c_j, c_{j-1}) + (1 - \exp(-d_j \rho_j)) \, \Pi_{c_{j-1}, c_j}, \quad (2)$$

where $\Pi$ is a transition matrix $\begin{pmatrix} \pi_0 & 1 - \pi_0 \\ 1 - \pi_1 & \pi_1 \end{pmatrix}$, $d_j$ is the distance between two adjacent SNPs at positions $(j-1)$ and $j$ on chromosome, and $\rho_j$ is the recombination rate for the same interval. The first term on the right-hand side of Equation (2) corresponds to the probability of no recombination events between the $(j-1)$th and the $j$th SNPs. On the other hand, the second term models a transition in the presence of a recombination event between the two SNPs. At recombination, the model can either transition to the same state, or to a different state. If the distance $d_j$ between the $(j-1)$th and $j$th SNPs is small or the recombination rate $\rho_j$ is low, the two SNPs are tightly linked, and it is likely that both SNPs will receive the same assignment for $c_{j-1}$ and $c_j$. Thus, the $c_j$'s for causal SNPs are set to 1 in a coupled manner, activating the corresponding covariates to take non-zero regression coefficients. We place a prior Beta($a_{00}, b_{00}$) on $\pi_0$, and Beta($a_{10}, b_{10}$) on $\pi_1$.

Our model differs from the fused lasso (Tibshirani et al. 2005) in that it encourages adjacent correlated covariates to take on the same assignment of whether they are relevant or not, while allowing each covariate to have its own regression coefficient. In the fused lasso, the adjacent regression coefficients themselves are encouraged to have the same values. In addition, our method directly makes use of the additional information in covariates such as the distance and recombination rate between two SNPs, whereas the fused lasso only uses the ordering information in covariates to fuse coefficients.

### 2.2 PARAMETER ESTIMATION

Because of the non-differentiability of the function used in Equation (1) for $\beta_j$ when $c_j = 0$, it is not possible to learn parameters of the model using an EM style algorithm commonly used for hidden Markov models. Instead, we use the Gibbs sampling to learn the parameters $\Theta = \{\boldsymbol{\beta}, \mathbf{c}, \sigma^2, \Pi\}$ of the model. In this section, we derive conditional posterior distributions of the parameters for Gibbs sampling.

For each of the $J$ covariates, we sample $\beta_j$ and $c_j$ from their joint posterior distribution

$$p(\beta_j, c_j|\boldsymbol{\beta}_{-j}, \mathbf{c}_{-j}, \mathbf{y}, \mathbf{X}, \sigma^2) = p(\beta_j|\boldsymbol{\beta}_{-j}, \mathbf{c}, \mathbf{y}, \mathbf{X}, \sigma^2)$$
$$P(c_j|\boldsymbol{\beta}_{-j}, \mathbf{c}_{-j}, \mathbf{y}, \mathbf{X}, \sigma^2). \quad (3)$$

We first sample $c_j$ from the marginal distribution, the second term on the right-hand side of Equation (3), after integrating out $\beta_j$. Conditional on the sampled $c_j$, we sample $\beta_j$ from its conditional posterior, the first term on the right-hand side of Equation (3).

In order to sample $c_j$, we re-write the second term on the right-hand side of Equation (3) as

$$P(c_j = k|\boldsymbol{\beta}_{-j}, \mathbf{c}_{-j}, \mathbf{y}, \mathbf{X}, \sigma^2)$$
$$\propto p(\mathbf{y}|\boldsymbol{\beta}_{-j}, c_j = k, \mathbf{X}, \sigma^2) P(c_j = k|c_{j-1}) P(c_{j+1}|c_j = k)$$

Sampling from the above equation requires to compute the marginal likelihood $p(\mathbf{y}|\boldsymbol{\beta}_{-j}, c_j, \mathbf{X}, \sigma^2)$ after integrating out $\beta_j$ when $c_j = 0$ and $c_j = 1$. When $c_j = 0$, the $j$th covariate is irrelevant, and we set $\beta_j = 0$. Thus, the marginal likelihood is simply given as

$$p(\mathbf{y}|\boldsymbol{\beta}_{-j}, c_j = 0, \mathbf{X}, \sigma^2)$$
$$= \left(\frac{1}{\sqrt{2\pi\sigma^2}}\right)^N \exp\left(-\frac{\sum_i (y_i - \mathbf{x}_i \boldsymbol{\beta})^2}{2\sigma^2}\right).$$

When $c_j = 1$, we compute the integral as below.

$$p(\mathbf{y}|\boldsymbol{\beta}_{-j}, c_j = 1, \mathbf{X}, \sigma^2)$$
$$= \int_{-\infty}^{\infty} \left(\frac{1}{\sqrt{2\pi\sigma^2}}\right)^N \exp\left(-\frac{\sum_i (y_i - \sum_k x_{ik}\beta_k)^2}{2\sigma^2}\right)$$
$$\cdot \frac{1}{2(2\lambda\sigma^2)} \exp\left(-\frac{|\beta_j|}{2\lambda\sigma^2}\right) d\beta_j$$
$$= K \int_{-\infty}^{\infty} \exp\left(-\frac{\sum_i (z_i - x_{ij}\beta_j)^2 + \frac{|\beta_j|}{\lambda}}{2\sigma^2}\right) d\beta_j$$
$$= K \left(\int_{-\infty}^{0} \exp\left(-\frac{\sum_i (z_i - x_{ij}\beta_j)^2 - \frac{\beta_j}{\lambda}}{2\sigma^2}\right) d\beta_j \right.$$
$$\left. + \int_{0}^{\infty} \exp\left(-\frac{\sum_i (z_i - x_{ij}\beta_j)^2 + \frac{\beta_j}{\lambda}}{2\sigma^2}\right) d\beta_j\right), \quad (4)$$

where $z_i = y_i - \sum_{k/j} x_{ik}\beta_k$, and $K = (1/2\pi\sigma^2)^{\frac{N}{2}}/(2(2\lambda\sigma^2))$.

Let $A_{(-)}$ denote the first integral and $A_{(+)}$ the second integral in Equation (4). Then, using a straightforward algebra, it can be shown that $A_{(-)}$ and $A_{(+)}$ are given as

$$A_{(-)} = \exp\left(-\left(\sum_i z_i^2 - \frac{(\sum_i z_i x_{ij} + 0.5/\lambda)^2}{\sum_i x_{ij}^2}\right)/(2\sigma^2)\right)$$
$$\cdot \sqrt{\frac{2\pi\sigma^2}{\sum_i x_{ij}^2}} \int_{-\infty}^{0} N_{(-)} \, d\beta_j$$

$$A_{(+)} = \exp\left(-\left(\sum_i z_i^2 - \frac{(\sum_i z_i x_{ij} - 0.5/\lambda)^2}{\sum_i x_{ij}^2}\right)/(2\sigma^2)\right)$$
$$\cdot \sqrt{\frac{2\pi\sigma^2}{\sum_i x_{ij}^2}} \int_{0}^{\infty} N_{(+)} \, d\beta_j,$$

where

$$N_{(-)} = N\left(\beta_j \Big| \frac{\sum_i z_i x_{ij} + \frac{1}{2\lambda}}{\sum_i x_{ij}^2}, \sigma^2 (\sum_i x_{ij}^2)^{-1}\right)$$
$$N_{(+)} = N\left(\beta_j \Big| \frac{\sum_i z_i x_{ij} - \frac{1}{2\lambda}}{\sum_i x_{ij}^2}, \sigma^2 (\sum_i x_{ij}^2)^{-1}\right).$$

Once we sample $c_j$ as described above, we sample $\beta_j$ from $p(\beta_j|\boldsymbol{\beta}_{-j}, \mathbf{c}, \mathbf{y}, \mathbf{X}, \sigma^2)$ in Equation (3). When $c_j = 0$, we set $\beta_j$ to 0. If $c_j = 1$, we re-write the conditional probability distribution as

$$p(\beta_j|\boldsymbol{\beta}_{-j}, \mathbf{c}, \mathbf{y}, \mathbf{X}, \sigma^2) = \frac{p(\mathbf{y}|\boldsymbol{\beta}, \mathbf{c}, \mathbf{X}, \sigma^2) p(\beta_j)}{\int p(\mathbf{y}|\boldsymbol{\beta}, \mathbf{c}, \mathbf{X}, \sigma^2) p(\beta_j) d\beta_j}. \quad (5)$$

We find that the denominator of Equation (5) is the same as what we computed in Equation (4). In fact, sampling from Equation (5) is equivalent to sampling from a mixture distribution of two components given as

$$m_j \sim \text{Bernoulli}\left(\frac{A_{(-)}}{A_{(-)} + A_{(+)}}\right)$$
$$\beta_j \sim \begin{cases} N_{(-),\beta_j<0} & \text{if } m_j = 0 \\ N_{(+),\beta_j>0} & \text{if } m_j = 1. \end{cases} \quad (6)$$

Using Equation (6), we augment $\beta_j$ with $m_j$, and sample $(\beta_j, m_j)$ by first drawing the mixture component label $m_j$ from the Bernoulli distribution and then drawing $\beta_j$ conditional on the $m_j$.

The conditional posterior for $\sigma^2$ is given as an inverse gamma distribution

$$\sigma^2|\boldsymbol{\beta}, \mathbf{c}, \mathbf{y}, \mathbf{X} \sim \text{Inv-gamma}((N + 2J + \nu_0)/2,$$
$$\big(\sum_i (\mathbf{y}_i - \mathbf{X}_i\boldsymbol{\beta})^2 + \frac{1}{\lambda}\sum_j |\beta_j| + \nu_0 s_0^2\big)/2\big).$$

Next, we sample the parameters $\pi_0$ and $\pi_1$ of the transition matrix $\Pi$. The conditional posterior for $\pi_0$ is

$$p(\pi_0|\mathbf{c}) \propto P(\mathbf{c}|\pi_0)p(\pi_0)$$
$$= \prod_{k \in S_{00}} (e^{-d_k\rho_k} + (1 - e^{-d_k\rho_k})\pi_0)$$
$$\cdot \prod_{k \in S_{01}} ((1 - e^{-d_k\rho_k})(1 - \pi_0))$$
$$\cdot \pi_0^{a_{00}-1}(1 - \pi_0)^{b_{00}-1}$$
$$\propto \left(\pi_0^{n_{00}+a_{00}-1}(1 - \pi_0)^{n_{01}+b_{00}-1}\right) \quad (7)$$

where $S_{ml} = \{k|c_{k-1} = m, c_k = l\}$ and $n_{ml}$ is the number of transitions from $c_{j-1} = m$ to $c_j = l$ in $\mathbf{c}$. We approximate $n_{00}$ as

$$n_{00} = \sum_j \frac{(1 - e^{-d_j\rho_j})\pi_0'}{e^{-d_j\rho_j} + (1 - e^{-d_j\rho_j})\pi_0'} I(c_{j-1} = 0, c_j = 0),$$

where $\pi_0'$ is the value from the previous sampling iteration, assuming that the number of events ($c_{j-1} = 0, c_j = 0$) due to $e^{-d_j\rho_j}$ and $(1 - e^{-d_j\rho_j})\pi_0$ are proportional to their probabilities. In Equation (7), the conditional posterior for $\pi_0$ is Beta($n_{00}+a_{00}, n_{01}+b_{00}$) for $\pi_0$. Similarly, we obtain the conditional posterior for $\pi_1$ as Beta($n_{11}+a_{10}, n_{10}+b_{10}$).

Finally, we sample $\lambda$ of the Laplacian prior from its conditional posterior

$$p(\lambda|\boldsymbol{\beta}, \sigma^2, \alpha, \gamma) = \text{Inv-gamma}\left(J' + \alpha, \frac{\sum_{j \in S_{J'}} |\beta_j|}{2\sigma^2} + \gamma\right),$$

where $J'$ is the number of covariates with $c_j = 1$ in the current sampling iteration, and $S_{J'}$ is the set of such covariates.

## 3 EXPERIMENTS

We demonstrate our proposed model on simulated data and mouse data, and compare the performance with those from the model with independent Bernoulli prior for $c_j$'s, ridge regression, and the lasso. We use ridge regression instead of ordinary least squares regression to prevent the singularity in matrix inversion, since often $J > N$ in our experiments. The regularization parameter in the ridge regression is set to a small value 0.1. The regularization parameter of the lasso is selected using a cross-validation.

In all of our experiments, for the block-regularized regression and the model with Bernoulli prior, we run the sampling algorithm for 5000 iterations after 2000 burn-in iterations. Samples are taken every 10 iterations. In the block-regularized regression, the priors for $\pi_0$ and $\pi_1$ are set to Beta(10,2), weakly encouraging the $c_j$'s to stay in the same state as the $c_{j-1}$'s. Similarly, in the model with Bernoulli prior, the prior for the parameter $p$ of the Benoulli distribution is set to Beta(10,2).

### 3.1 SIMULATION

We generate 360 haplotypes of a 40kb region with mutation rate 0.8/kb and recombination rate 0.1/kb using the software *ms* (Hudson 2002), and retain only those SNPs whose minor allele frequency is greater than 0.01. In the haplotypes generated under this setting, there are 108 SNPs in the region and 23 blocks in which no recombination events occurred. Then, we randomly pair two haplotypes to obtain genotypes of 180 individuals. We run the widely used software *Phase 2.1.1* (Li and Stephens 2003) on these data to estimate the recombination rate $\rho$ between each pair of adjacent SNPs. We select 10 SNPs as relevant variables that are grouped into three blocks of 3, 2, and 5 SNPs respectively such that the SNPs within a group are located within a block of no recombination given by *ms*. Based on these selected causal SNPs, we generate a phenotype for each individual with $\beta_j = 2.5$ for causal SNPs and $\beta_j = 0$ for non-causal SNPs. The random noise generated from $N(0,1)$ is added to the simulated phenotype. The true parameters used in this simulation are shown in Figure 2(a). The dots show values for $\beta_j$'s, and the line indicates the locations of the causal SNPs, where 1's represent causal SNPs, and 0's non-causal SNPs.

Using the data simulated from the true parameters in Figure 2(a), we fit our proposed model, the model with independent Bernoulli prior, ridge regression and the lasso, and plot the estimated $\boldsymbol{\beta}$ in Figures 2(b)-(e). For the block-regularized regression and the model with independent Bernoulli prior, we show the sam-ple corresponding to the lowest train error, and plot the estimated **c** for the same sample as a line. The block-regularized regression in Figure 2(b) discovers the block structure in covariates with four groups of relevant variables, three of which roughly correspond to the three groups in the true parameters. Throughout our simulation experiments, we found that the groups of causal SNPs indicated in $c_j$'s estimated by the block-regularized regression tend to extend to a slightly larger interval than in the true parameters, and that a subset of such SNPs with $c_j = 1$ has a large value for $\beta_j$. Thus, we can view $c_j$'s as suggesting regions of relevant variables, and $\beta_j$'s as deciding how relevant the variables with $c_j = 1$ are. As we see in Figures 2(c)-(e), the block structure is not obvious in the results from the other three methods.

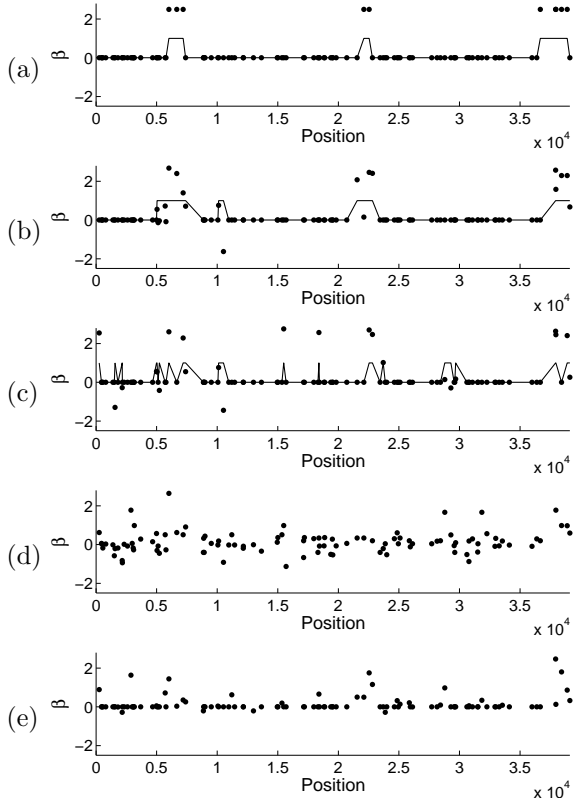

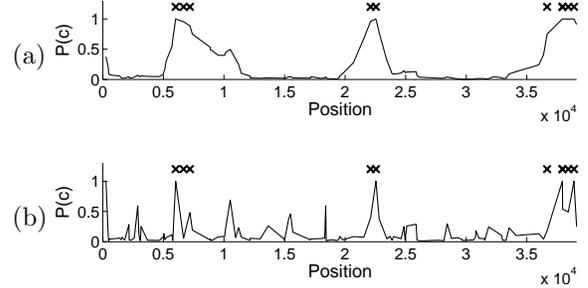

Figure 3: Estimated $P(c_j)$'s for (a) block-regularized regression and (b) the model with independent Bernoulli prior, corresponding to the results in Figure 2(b) and (c) respectively. The ×'s indicate the locations of true relevant variables.

Table 1: Summary Statistics of Simulated Data

| $\rho$ | Number of SNPs | | | Average number of SNPs per block |
|---|---|---|---|---|
| | Min | Max | Mean | |
| 0.05/kb | 86 | 260 | 150.8 | 5.59 |
| 0.1/kb | 103 | 226 | 147.2 | 5.53 |
| 0.5/kb | 99 | 214 | 151.0 | 1.18 |
| 1.0/kb | 110 | 197 | 152.2 | 0.57 |

Figure 2: Simulation results with $\rho$=0.1/kb and $\beta_j = 2.5$ for relevant variables. True parameters are shown in (a), and the estimated parameters are shown for (b) the block-regularized regression, (c) the model with independent Bernoulli prior, (d) ridge regression, and (e) the lasso. Dots indicate $\beta_j$'s at each position, and the lines in (a), (b), and (c) represent $c_j$'s.

Figures 3(a) and (b) show the estimated $P(c_j)$'s for the block-regularized regression and the model with independent Bernoulli prior respectively, using the same data as in Figure 2. Each $P(c_j)$ is estimated as the proportion of the number of times that the $c_j$ is set to 1 in samples for the $c_j$. The locations of the true causal SNPs are marked with ×'s. The block-regularized regression encourages a block structure among relevant and irrelevant covariates, leading to a smoother variation in $P(c_j)$'s between adjacent covariates compared to the model with independent Bernoulli prior.

In order to quantify the performance of the block-regularized regression and various other regression methods in the presence of different level of correlation among SNPs, we repeat the above procedure of simulating data for four different recombination rates $\rho$=0.05/kb, 0.1/kb, 0.5/kb, and 1/kb, using $\beta_j = 2.0$ for causal SNPs and $\beta_j = 0$ for non-causal SNPs. A lower recombination rate results in a more tightly linked SNPs and a stronger block structure in covariates. For each recombination rate, we generate 50 datasets of 180 individuals, and report the results averaged over these 50 datasets for the given recombination rate. Because of the random nature in the data generation process of the simulation software *ms*, the number of SNPs vary in each dataset even if the same parameters are used in simulation. The minimum, maximum, and average number of covariates in the 50 datasets for each recombination rate are shown in Table 1 as well as the average number of SNPs within a block given by *ms* during simulation. As the recombination rate gets higher, the correlations between adjacent SNPs become weaker, leading to only less than 1 SNP per block in the case of $\rho = 1.0$/kb.

Given these datasets generated as described above, we estimate $\boldsymbol{\beta}$ with the block-regularized regression, the model with independent Bernoulli prior, ridge regression, and the lasso, and plot precision-recall graphs in Figure 4(a)-(d) for recombination rates $\rho=0.05$/kb, 0.1/kb, 0.5/kb, and 1.0/kb respectively. To obtain each precision-recall curve in Figure 4, we estimate $\beta_j$'s given a dataset and a regression method, and rank the SNPs according to the absolute values of $\beta_j$'s. The SNP with the largest value of $|\beta_j|$ is considered as the most relevant. We compare the rankings of SNPs given by each regression method to the list of true causal SNPs, and compute the precisions and recalls shown in Figure 4. As can be seen in Figures 4(a) and (b), the block-regularized regression clearly outperforms other methods, since the correlations among SNPs are relatively high, and the block-regularized regression takes advantage of this correlation structure. As the recombination rate increases in Figures 4(c) and (d), the advantage of having a block model decreases. When $\rho=1$/kb, the average number of SNPs per block is less than 1 as shown in Table 1, and the block-regularized regression, the model with independent Bernoulli prior and the lasso perform similarly. We can use $P(c_j)$'s instead of $|\beta_j|$'s to rank the SNPs. When we plotted the precision-recall graphs according to the $P(c_j)$'s, we obtained similar results to the ones in Figure 4.

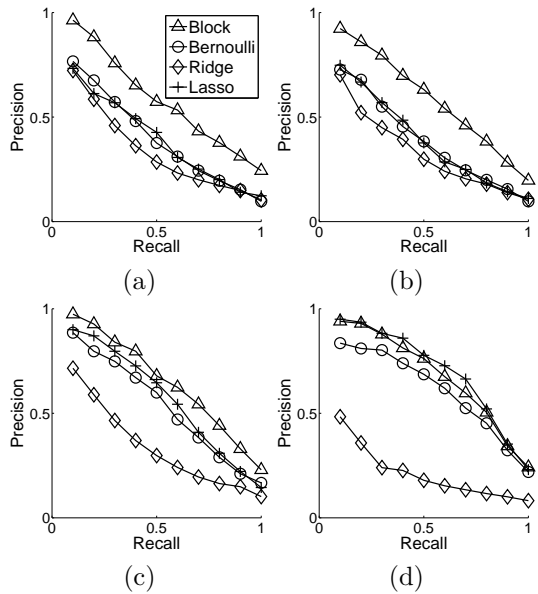
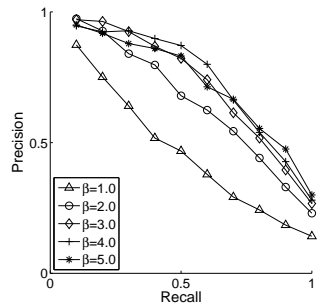

Figure 4: Precision-recall graphs for simulated data. (a) $\rho = 0.05$, (b) $\rho=0.1$, (c) $\rho=0.5$, and (d) $\rho=1.0$/kb.

In Figure 5, we fit the block-regularized regression model to the dataset simulated with different values of $\beta_j$ for relevant variables and the noise distributed as $N(0,1)$, and plot the precision-recall graphs. The recombination rate $\rho$ is set to 0.5/kb, and each precision-recall curve is the result averaged over 50 datasets. We see that as the signal to noise ratio increases, the performance increases.

Figure 5: Precision-recall graphs for block-regularized regression, using simulated data with varying $\beta_j$'s for relevant variables.

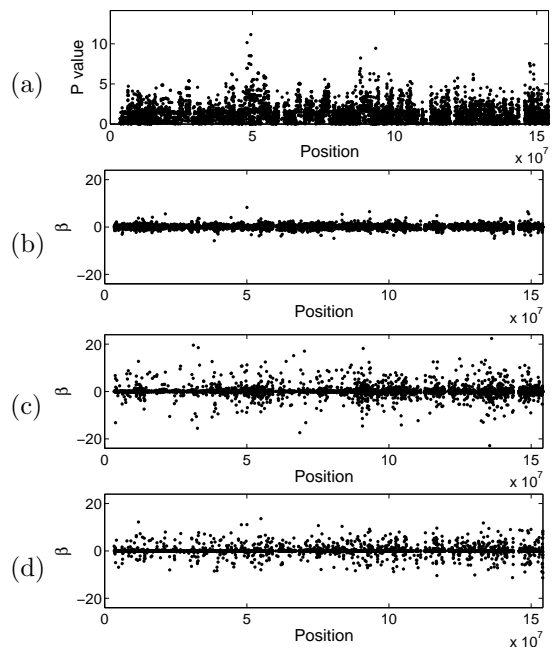

Figure 6: Results for the mouse haplotype data (chromosome 4) and the measurements of drinking preference. (a) $-\log(p$ value), (b) block-regularized regression, (c) the model with independent Bernoulli prior, (d) the lasso.

### 3.2 MOUSE DATA

We apply the block-regularized regression to the inbred laboratory mouse haplotype map publicly available from BROAD institue website[1]. We use the measurements for the sodium intake of 25% NaCl concentration for female mice (Tordoff et al. 2007) as phenotype. The data for 33 strains are available for both the genotype and phenotype dataset. Thus, the number of individuals in this experiment is 33. We consider the 8217 SNPs in chromosome 4 of length 154Mb as covariates. We are interested in scanning the chromosome and discovering the SNPs with high genetic effects on the phenotype.

---
[1] http://www.broad.mit.edu/mouse/hapmap

The most commonly used method for discovering SNPs highly associated with the phenotype is to perform a statistical test for the phenotype and one SNP at a time and report the SNPs with high $p$-values as significant. The result for the mouse data using one such test, the Wald test, is shown in Figure 6(a). The $y$-axis shows $-\log(p\text{-value})$ for each SNP. In Figure 6(b)-(d), we show the estimated $\boldsymbol{\beta}$ using the block-regularized regression, the model with independent Bernoulli prior, and the lasso respectively. For these three regression models, we divide the whole sequence into segments of 200 SNPs, and fit the model to one segment at a time. We see that the SNPs with high values of $-\log(p\text{-value})$ in Figure 6(a) roughly correspond to the SNPs with high $\beta_j$ values in Figure 6(b).

## 4 CONCLUSIONS

In this paper, we considered the problem of finding a subset of covariates in a high-dimensional space that affect the output variable when there is a block structure in the covariates. In the context of association mapping, we proposed a regression-based model with a Markov chain prior that encodes the information in the correlation structure such as distance and recombination rate between adjacent SNP markers. We demonstrated on the simulated and mouse data that our proposed algorithm can be used to identify groups of SNP markers as a relevant block of causal SNPs.

The idea of representing the correlation structure as a Markov chain in a variable selection method to learn grouped relevant variables can be generalized to use a graphical model as a prior in a variable selection problem to represent an arbitrary correlation structure in variables in a high-dimensional space. Another interesting extension of the model is to model a structure in output variables as well when measurements of multiple output variables are available.

**Acknowledgements**

This research was supported by NSF Grants CCF-0523757 and DBI-0546594.